\documentclass[twocolumn,superscriptaddress]{revtex4}

\usepackage[dvips]{graphicx}
\usepackage{amssymb,amsfonts,amsmath}

\newcommand\sfrac[2]{{\textstyle{\frac{#1}{#2}}}}

\newcommand{\bs}{{\bf {s}}}
\newcommand{\br}{{\bf {r}}}
\newcommand{\bv}{{\bf {v}}}
\newcommand{\bk}{{\bf {k}}}
\newcommand{\bom}{{\mbox{\boldmath $\omega$}}}

\begin{document}

\title{Coherent vortex structures in quantum turbulence}

\author{A.~W.~Baggaley}
\email{a.w.baggaley@ncl.ac.uk}
\author{C.~F.~Barenghi} 
\author{A.~Shukurov} 
\affiliation{School of Mathematics and Statistics, University of
Newcastle, Newcastle upon Tyne, NE1 7RU, UK}
\author{Y.~A.~Sergeev} 
\affiliation{School of Mechanical and Systems Engineering, Newcastle
University, Newcastle upon Tyne, NE1 7RU, UK}

\begin{abstract}
This report addresses an important question discussed by the quantum
turbulence community during the last decade: do quantized vortices
form, in  zero-temperature superfluids, coherent structures
similar to vortex tubes in ordinary, viscous turbulence? So far
the evidence provided by numerical simulations is that bundles of
quantized vortices appear in finite-temperature superfluids,
but from the interaction with existing coherent structures in the
turbulent (viscous) normal fluid, rather than due to the intrinsic superfuid
dynamics. In this report we show that, in very intense quantum turbulence (whose simulation was
made possible by a tree algorithm), the vortex tangle contains
small coherent vortical structures (bundles of quantized vortices)
which arise from the Biot--Savart dynamics alone, and which are similar to the
coherent structures observed in classical viscous turbulence.
\end{abstract}


\maketitle

It has been known since the 1980's
that homogeneous isotropic turbulence
contains intermittent worm-like regions of concentrated vorticity \cite{Siggia,Kerr,Sreeni85,Douady,Moffatt,Sreeni-Antonia,Moisy-Jimenez}.
Their role in the dynamics of turbulence is not clear
\cite{Frisch}: there is no consensus as to whether they are
responsible for the main properties of turbulence (e.g the
celebrated Kolmogorov energy spectrum) or only affect the tails of
statistical distributions and the exponents of high-order
structure functions. Turbulence is also studied at temperatures
near absolute zero in superfluid helium
\cite{Vinen,Manchester,Lancaster,Helsinki}. Here the viscosity is
zero, and the vorticity is concentrated in thin, discrete vortex
lines of fixed core radius and circulation (whereas vortices in ordinary
fluids are viscous and can have any size and strength). This
`quantum turbulence' shares important properties with the ordinary
(classical) turbulence~\cite{Vinen,Skrbek,Lvov_2006},
from the same drag force on a moving
sphere~\cite{VanSciver} to the Kolmogorov energy spectrum in homogeneous
isotropic turbulence
\cite{Tabeling,Grenoble,Nore,Tsubota-vd,Tsubota-GP,Baggaley-density}. Here we show
that, alongside the Kolmogorov spectrum, quantized vortices tend to
form coherent structures, similar to the vortex tubes of ordinary
turbulence. Despite its relative simplicity, quantum turbulence seems to
share important features of generic turbulent flows.

At temperatures $T$ below $0.7\,\rm K$, thermal excitations become
ballistic (in that sense that their propagation is not affected by any interaction with
the background) so that liquid $^4$He is effectively a pure superfluid.
Because of the quantum-mechanical constraints on the rotation, its
flow is everywhere potential with the exception of vortex line
singularities of atomic thickness (the core radius is
approximately $a_0 \approx0.1\,\rm nm$), each line carrying one
quantum of circulation $\kappa=h/m=10^{-7}\,\rm m^2/s$, where $h$
is Planck's constant and $m$ is the mass of the single helium atom. (In
liquid $^3$He-B, which becomes superfluid at much lower
temperatures, $T<1\,{\rm mK}$, the core radius is about
$80\,{\rm nm}$, and the quantum of circulation is
$\kappa=h/(2m)=0.662\times10^{-7}\,\rm m^2/s$, where $m$ is the atomic
mass of the $^3$He isotope, whose nuclei are fermions.) Quantum turbulence,
readily created by
stirring the liquid helium, takes the form of an apparently
disordered tangle of such vortex lines. The question which we ask
is whether discrete quantized vortex filaments organize themselves
and form coherent structures, as in ordinary turbulence.

To model quantum turbulence in the zero-temperature limit, we
observe that $a_0$ is many orders of magnitude smaller than the
typical distance, $\ell \approx 10^{-4}$--$10^{-6}\,\rm m$,
between vortex lines in experiments, so it is appropriate to
describe vortex lines as space curves ${\bf s}(\xi,t)$ of
infinitesimal thickness which move according to the Biot--Savart
law~\cite{Saffman}
\begin{equation}
\frac{d{\bf s}}{dt}=-\frac{\kappa}{4 \pi} \oint_{\cal L}
\frac{({\bf s}-{\bf r}) } {\vert {\bf s} - {\bf r} \vert^3} \times
{\bf dr}, \label{eq1}
\end{equation}
where ${\bf s}$ is the position vector of a point on the vortex line,
$t$ is time, $\xi$ is the arc length measured along the vortex line, and the
integral extends over the entire vortex configuration $\cal L$.
Equation~(\ref{eq1}) states classical incompressible
Euler dynamics in integral
form. However, unlike classical Euler vortices, quantized vortices
reconnect if they come sufficiently close to each other, as seen
directly in recent experiments \cite{Maryland} and demonstrated
using the Gross--Pitaevskii equation for the Bose--Einstein
condensate~\cite{Koplik}. To apply equation~(\ref{eq1}) to
superfluid helium we must therefore add an algorithmic
reconnection procedure~\cite{Schwarz}.

\begin{figure}[h]
\begin{center}
\includegraphics[width=0.4\textwidth]{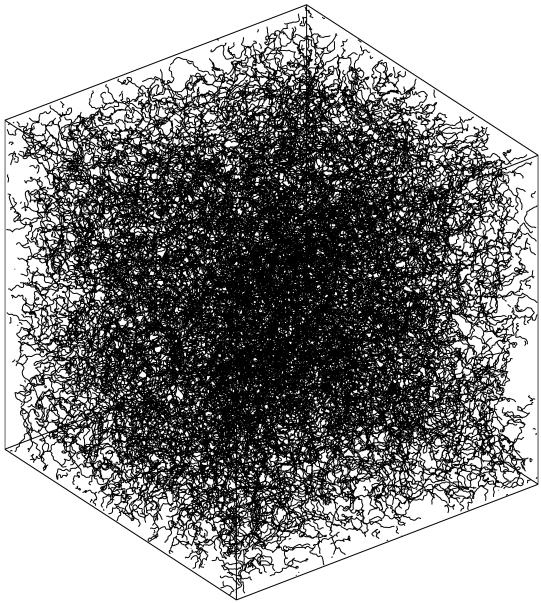}
\caption{The instantaneous form of the saturated vortex tangle at
$t=0.06\,\rm s$.}
\label{fig1}
\end{center}
\end{figure}

We perform numerical simulations of reconnecting vortex lines at
$T=0$ evolving according equation (\ref{eq1}) in a period cubic
box of the size $D=0.75\times10^{-3}\,\rm m$. The technique to discretize the
vortex lines, the regularization of the Biot--Savart integral,
and the vortex reconnection procedure are standard and well described in the
literature~\cite{Schwarz,Tsubota-vd}. Our numerical algorithm,
which controls the discretization, enforces a separation $\Delta
\xi$ between points on the vortex line to remain between $2 \times 10^{-6}\,\rm m$
and $4 \times 10^{-6}\,\rm m$ during the evolution; this is
therefore the numerical resolution of the computation. The time
integration uses a third-order Adams--Bashforth scheme; the
typical time step is $\Delta t=4 \times 10^{-6}\,\rm s$. Our vortex
reconnection procedure guarantees that a small amount of vortex
length (as a proxy for kinetic energy) is lost at each reconnection, in
agreement with numerical calculations of vortex reconnections
performed using the Gross--Pitaevskii equation \cite{Leadbeater}.
The reconnection procedure and the enforcement of the minimum
distance $\Delta \xi$ provide the numerical dissipation mechanism which
plays the r\^ole of phonon emission in actual $^4$He \cite{Baggaley-cascade}
(or of the Caroli--Matricon~\cite{Kopnin} mechanism in $^3$He-B). 
To cope with the
large number $N\sim 10^5$ of discretization points along the
vortex lines, we use a 
tree algorithm \cite{Barnes,Baggaley-density,Baggaley-long} which reduces the
CPU time needed to evaluate the Biot--Savart integrals from $N^2$ to
$N\log{N}$. The initial condition consists of randomly oriented straight
vortex lines, each carrying one quantum of circulation
(however, we checked that results do not depend on
the initial configuration of vortices). We find that the initial
vortices interact and reconnect, and, after a transient of the
order of $0.03\,\rm s$, the vortex line density (vortex length per
unit volume) saturates at the average value $L=7.9 \times 10^8\,\rm
m^{-2}$. The average curvature saturates too. The
vortex tangle is shown in Figure~\ref{fig1}. By construction, the circulation
of each vortex line is conserved, 
and each vortex moves exclusively due to the collective
influences of all the other vortex lines; this represents an accurate model of quantum
turbulence.

\begin{figure}[h]
\begin{center}
\includegraphics[width=0.4\textwidth]{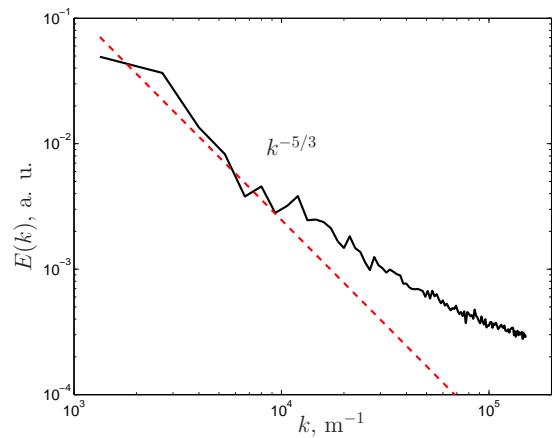}
\caption{The energy spectrum $E(k)$ for the flow of
Figure~\ref{fig1}. The red dashed line shows the classical Kolmogorov
scaling $k^{-5/3}$. 
}
\label{fig2}
\end{center}
\end{figure}

One of the most important quantity in theory of
turbulence is the energy spectrum, $E(k)$ defined by
\begin{equation}
E=\frac{1}{V} \int_V \sfrac{1}{2} \vert\bv\vert^2\, dV
        =\int_0^{\infty} E(k)\,dk\,,
\label{spectrum}
\end{equation}
where $E$ is the kinetic energy (per unit mass), $V$ is volume,
$k=\vert\bk\vert$, with $\bk$ the wave vector. 
Figure~\ref{fig2} shows a spectrum typical of quantum turbulence.
The Biot--Savart
interactions of vortex lines over length scales larger than the
average inter-vortex distance $\ell \approx L^{-1/2}$ has induced
the same Kolmogorov energy spectrum $E(k)\sim k^{-5/3}$ (for
$k \ll k_{\ell}=2 \pi/{\ell}$) which is observed in ordinary turbulence
\footnote{
The wavenumber corresponding to the
average inter-vortex spacing in Figure~\ref{fig2}
is $k_{\ell}=1.8 \times 10^5\,\rm m^{-1}$,
where the intervortex distance is estimated in the usual
way as $\ell\approx L^{-1/2}$. Note that in our calculation, due
to the the absence of friction, 
the vortex
filaments are very wiggly, so the use of the actual filaments' length
(rather than a suitably defined smoothed length) overestimates $k_{\ell}$.
}.
The Kolmogorov spectrum was also observed in experiments with
turbulent superfluid helium \cite{Tabeling,Grenoble}, and in calculations
performed with both the vortex filament model
\cite{Tsubota-vd,Baggaley-density} and the Gross--Pitaevskii equation
\cite{Nore,Tsubota-GP}. At larger wavenumbers 
the spectrum becomes shallower due to the Kelvin-wave cascade along
individual vortex lines \cite{Baggaley-cascade,Kozik,Lvov}, in agreement
with Refs. \cite{Tsubota-vd,Sasa}. The Kelvin
cascade (which is not the main concern in this work) is due to the nonlinear
interaction of  Kelvin waves (helical perturbations of vortex filaments). It
transfers kinetic energy downscale, creating shorter and shorter
waves, until, at very short length scales, phonon emission turns
the kinetic energy of the waves into the sound energy. (In $^3$He-B,
the dissipation in the zero-temperature limit is likely to be
associated with a different, Caroli-Matricon mechanism of energy
loss from Kelvin waves into the quasiparticle bound states.) In
our calculation, a small numerical dissipation arising from the
discretization and the vortex reconnection procedure plays the
role of phonon emission \cite{Baggaley-cascade}. If we continued the
calculation for a longer time, this numerical dissipation would
make the vortex length to decay, but the energy and the length can be considered
effectively constant over the timescale of
interest here.

\begin{figure}[h]
\begin{center}
\includegraphics[width=0.4\textwidth]{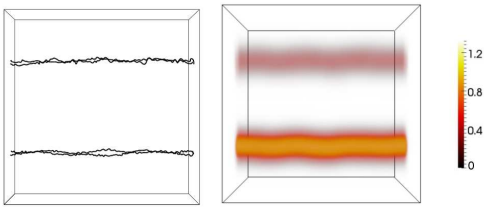}
\caption{ Volume rendering (a semi-transparent representation)
of the magnitude of the smoothed
vorticity field $\vert \bom(\br) \vert,\,{\rm s}^{-1}$, from
antiparallel (top) and parallel (bottom) vortex pairs. In the
former, the average vorticity is small but not zero due to the presence of
small-amplitude Kelvin waves (note the wigglyness of the vortex
filaments). In the latter, the contributions of the two vortex strands
add up.}
\label{fig3}
\end{center}
\end{figure}

To highlight the presence of any structure in the vortex tangle
shown in Figure~\ref{fig1}, we convolve our discrete vortex
filaments with a Gaussian kernel, and define a smoothed vorticity
field $\bom$ as
\begin{equation}
\bom(\br,\,t)=\kappa\sum_{i=1}^N
\frac{\bs'_i}{(2\pi\sigma^2)^{3/2}}
\exp(-\vert\bs_i-\br\vert^2/2\sigma^2)\Delta\xi\,, \label{eq:smooth}
\end{equation}
where $\bs'_i=d\bs_i/d\xi$ is the unit vector along a vortex at
the discretization point $\bs_i=\bs_i(\xi,t)$, and the smoothing
length $\sigma$ is of the order of $\ell$. We test that, under
this smoothing operation, a collection of randomly oriented vortex
lines whose separation is of the order of $\ell$ yields
$\bom\approx{\bf 0}$; conversely, an organized bundle of vortex
filaments of the same circulation yields a smooth vorticity distribution.
Because of the wigglyness of the vortex lines at short
lengthscales created by the Kelvin cascade, we also test the
smoothing procedure on vortex lines after adding ten helical waves
with an imposed $k^{-7/5}$ Kelvin-wave amplitude spectrum, as shown in Figure~\ref{fig3}.

\begin{figure}[h]
\begin{center}
\includegraphics[width=0.4\textwidth]{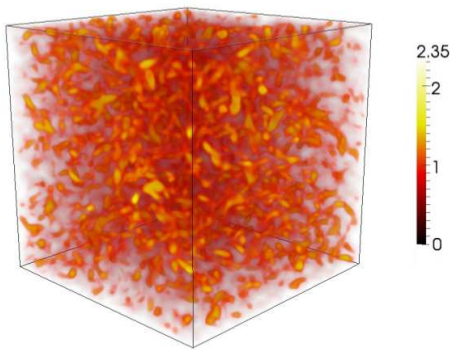}
\caption{ Volume rendering of the magnitude of the smoothed
vorticity field of Figure~\ref{fig1}, $\vert\bom(\br)\vert,\,{\rm s}^{-1}$.
Note the vortical worm-like structures, similar
to the coherent structures observed in ordinary turbulence.}
\label{fig4}
\end{center}
\end{figure}

Figure~\ref{fig4} shows the smoothed vorticity corresponding to
Figure~\ref{fig1}: vortical `worms', such as those seen in
numerical simulations of ordinary turbulence, are clearly visible.

This is clear evidence of three-dimensional coherent
superfluid vortex structures at $T=0$ which arise simply by the fundamental
equation of motion~(\ref{eq1}) of vortex dynamics.
There have been reports on the existence of
vortex bundles \cite{Kivotides,Morris}
in Biot--Savart simulations of quantum turbulence at relatively
high temperatures,
$0.7\,{\rm K}<T<T_\lambda\approx2.17\,{\rm K}$,
where thermal excitations form a viscous normal fluid component which
interacts with the vortex lines via the mutual friction force. However, in this
finite-temperature regime, vortical structures belong to the normal turbulent
fluid, and the friction force
\cite{Hulton} would naturally induce superfluid bundles around them
(although their stability is an open question). Here we focus on the
development of vortical bundles in the perfect superfluid, as a 
consequence of pure incompressible Euler dynamics. 
An intermittent vortex structure has also been noticed in a recent \cite{Sasa}
Gross--Pitaevskii simulation of turbulence in a Bose--Einstein condensate.
However, in their calculation the
vortex cores were
(typically of a Bose--Einstein condensate, either realized numerically 
or in the laboratory) very close to each other:
$\ell/a_0 \approx 4.5$ to $9$, in contrast to our much larger 
$\ell/a_0 \approx 3.5 \times 10^3$ typical of $^4$He and $^3$He-B superfluids;
compressible effects 
(rapid density changes near vortex cores, sound waves, and the application of
artificial damping which affects vortex positions)
which are absent in our (incompressible) Biot--Savart model were also
likely to have played a r\^ole. 

\begin{figure}[h]
\begin{center}
\includegraphics[width=0.38\textwidth]{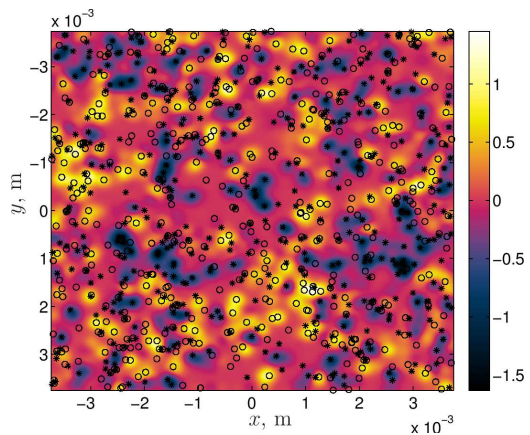}
\caption{The cross-section $z=0$ of the $z$-component of the smoothed
vorticity field from Figure~\ref{fig4}, $\omega_z(\br),{\rm s}^{-1}$. Open
circles show where vortex lines cross the plane from $z<0$,
and asterisks show those coming from $z>0$. The clustering of vortex points of the same
sign is evident, visible as regions of large positive and negative
smoothed values of $\omega_z$ (shown blue and yellow). }
\label{fig5}
\end{center}
\end{figure}

To verify that the structures in Figure~\ref{fig4}
are indeed bundles of quantized
vortices, we show in Figure~\ref{fig5} a two-dimensional cross-section of the
$z$-component of the smoothed vorticity $\bom$
in the $z=0$ plane of Figure~\ref{fig4}. We overlay
the intersections of vortex filaments with the plane $z=0$
distinguishing the sign of $\omega_z$.
It is apparent that the smoothed vortical structures
shown in Figure~\ref{fig4} are indeed small bundles
of aligned vortex filaments, which appear as small clusters of
positive and negative vorticity.
There are typically 2 to 5 vortex points in each cluster.

\begin{figure}
\begin{center}
\includegraphics[width=0.34\textwidth]{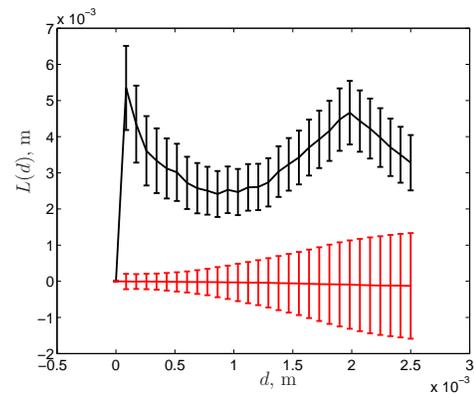}
\caption{Besag's function, $L(d)$, averaged over a random selection of
1,000 
two-dimensional cross-sections (parallel to one of the coordinate
planes) of the vortex tangle of Figure~\ref{fig4}
in the final, statistically steady state (black) and the  initial state (red).
Vertical bars show the one standard deviation range. Note that $L$ increases
with $d$ sharply from $L=0$ at $d=0$ to its maximum value for
$d\approx4\times10^{-6}\,{\rm m}$, and then varies smoothly with $d$,
reaching the secondary maximum at $d\approx2\times10^{-3}\,\rm m$. The fact that in the final state$L(d)$
is significantly larger than zero, in
the whole range of $d$ explored, confirms the existence of coherent structures
(clusters of vortex filaments in each cross-section) of various scales $d$, with
some predominance of $d\approx2\times10^{-3}\,\rm m$. On the contrary, the initial
state has $L=0$ within errors at all scales, consistently with the purely random nature
of the initial conditions.}
\label{fig6}
\end{center}
\end{figure}

It is important to check that the vortex bundles
are physically distinct coherent structures of a well-defined scale,
 rather than just a part of a
purely random distribution, which would also contain structures of any scale
which would become prominent after averaging. Clustering in
any two-dimensional system of points (or of any other discrete objects)
can be confirmed and quantified using Ripley's $K$-function~\cite{Ripley}, or,
alternatively, Besag's $L$-function~\cite{Besag}. These functions,
frequently used in applied statistics, were also used to detect spatial
correlations in two-dimensional physical systems (in particular in
Monte-Carlo simulations and also in biological applications of the Ising model).
Ripley's $K$-function is defined as
\begin{equation}
K(d)=\left(\frac{D}{M}\right)^2 \sum_{i=1}^M \sum_{j=1, j\ne i}^M I(d_{i,j}<d)\,,
\label{RipleyK}
\end{equation}
where $M$ is the number of points within the area $A=D^2$ (with $D$ the
size of the domain),
$d_{i,j}$ is the distance between points $i$ and $j$, and
the $I(x)$ is unity if the condition $x$ is satisfied and zero
otherwise. However, the related Besag's function is more convenient for our
purposes, defined as
\begin{equation}
L(d)=\sqrt{K(d)/\pi}-d\,;
\label{BesagL}
\end{equation}
$L=0$ means complete spatial randomness,
$L<0$ implies dispersion, and $L>0$ aggregation (clustering).
We show in  Figure~\ref{fig6} Besag's function for a large collection of two-dimensional
cross-sections of Figure~\ref{fig4}.  The result confirms
that vortex points of the same sign tend to cluster, hence the
vortex bundles of Figure~\ref{fig4} represent distinct physical entities
rather than merely an element of a purely random distribution of vorticity.

We now proceed with the interpretation of our results, paying
particular attention to the relation to the ordinary turbulence.
According to our calculations, parallel vortex lines tend to come together
locally, forming clusters of vortex filaments of the same circulation
in any cross-section through the flow. If the vorticity distribution was
two-dimensional, the well-known inverse enstrophy cascade of the two-dimensional
turbulence would transfer vorticity to large scales, creating large-scale
regions populated by vortices of the same
sign. The two-dimensional inverse cascade is due to the conservation of
the enstrophy, $Q=\int_V \vert\bom\vert^2\, dV$,
the second invariant of the Euler equation in two
dimensions. In three dimensions, only the first integral exists,
the kinetic energy $E=\int_V \sfrac12 \vert\bv\vert^2\, dV$, whose conservation
results in the direct cascade of the energy to smaller scales in both two-
and three-dimensional turbulent flows.

To find whether there is a link between formation of three-dimensional
coherent structures and tendency of vortices to cluster in two dimensions
we performed additional numerical experiments.
The two-dimensional system of interacting point vortices in the inviscid
fluid was introduced by Onsager~\cite{Onsager} in 1949, and is known
as the Onsager point-vortex gas.
This is where the concept of inverse cascade was born.
Onsager found that, statistically, the states of
the two-dimensional system corresponding to the formation of large-scale
coherent structures (clusters of point vortices of the same polarity)
formally have negative temperature (a detailed analysis was later provided by
Montgomery~\cite{Montgomery}). The ordering of vortices and the
properties of the inverse cascade in the point-vortex gas have been extensively studied,
with the numerical study of the formation of large-scale circulation
patterns in the two-dimensional periodic domain by Montgomery
and Joyce~\cite{Montgomery-Joyce} being one of the first studies of this kind. Recently some of the
authors of the present report demonstrated~\cite{Wang} the clustering of point
vortices of the same polarity in a small system of just a few hundred
vortices, by computing the dis balance between the vortices and
antivortices within subdomains of the periodic two-dimensional computational
domain. Here we have simulated the evolution of a much
larger number (10,000) of point vortices in two dimensions.
The system evolves from a statistically uniform state with
zero net circulation where the number of vortices equals that of
antivortices. Using Besag's function, we have confirmed quantitatively the
non-random clustering of vortex points.

To check if coherent structures discussed above could be
a fossil of the two-dimensional reverse cascade that survives in three
dimensions,
we explored a modified two-dimensional Onsager's system of point vortices
similar to that shown in Figure~\ref{fig5}. We modelled the vortex motion out of the plane
by removing randomly chosen vortex-antivortex pairs from the plane at a certain rate, and then randomly
re-introducing them at random positions elsewhere in the plane.
For example, if a moving vortex loop in three dimensions crosses the plane,
the number of point vortices in the plane first increases by two, and then is reduced by two
as the loop leaves the plane.
As these disturbances violate the local enstrophy conservation, they can be used
to model, in a heuristic manner, local three-dimensional effects.
We found that Besag's function in such a disturbed two-dimensional system
still shows tendency to clustering typical of the inverse enstrophy cascade,
but weaker than in the pure two-dimensional case. The (positive) value of
Besag's function depends on the rate of vortex removal.
It is therefore likely that the two-dimensional tendency to cluster
is limited by the fact that the enstrophy is not constant, but this
tendency is not entirely suppressed entirely.

When interpreting our results, it is also useful to compare them with
recent calculations of Numasato et al.~\cite{Numasato}, who
observed only the direct energy cascade in a
two-dimensional Bose--Einstein condensate. These authors
solved the Gross--Pitaevskii equation; this model
allows vortex reconnections, hence does not conserve the
number of vortices (i.e., the enstrophy).  These authors also argued that the
Gross--Pitaevskii model is
compressible (the condensate's density vanishes at the vortex axes and
large-scale sound waves are easily excited, as noticed by White
et al.~\cite{White}), and the compressibility alone
could explain the lack of the inverse cascade.
Can compressibility be equally important in our case? 
As we already noted above, in 
a typical atomic Bose--Einstein condensate, either realized numerically
or in the laboratory, the distance between the vortices, $\ell$, is only a few
times larger than the vortex core radius $a_0$, while $\ell/a_0 =O(10^4)$
in the superfluid helium. Therefore, compressibility is less likely to prevent
the formation of coherent structures in the superfluid turbulence.

In the ordinary turbulence, the origin of coherent vorticity structures is
often attributed to the roll-up of vortex sheets by the Kelvin--Helmholtz
instability~\cite{Vincent-Meneguzzi1994}. The superfluid turbulence considered
here has no vortex sheets, so the coherent structures observed here seem to
involve a different mechanism.

Our future work will explore if the number of vortex lines in the bundles
increases as the circulation is reduced (with kinetic energy fixed), as one
expects that the limit $h \to 0$ should yield the classical behaviour where
the range of vorticity values in coherent structures is wider than that
observed in our numerical experiments. To better understand the dynamics of
coherent structures, we also plan to investigate the time evolution of
Ripley's and Besag's functions for the three-dimensional vortex tangle and to
explore the time correlations of the flow.

In summary, our numerical experiments with very intense quantum turbulence at
absolute zero have revealed that the vortex tangle contains coherent vortical
structures, or bundles of vortex lines, which arise from the Biot--Savart
dynamics alone, and appear to be similar to the vorticity `worms' observed in
the ordinary turbulence. This result sheds new light on the relation between
the ordinary and quantum turbulent flows, suggesting that their connection can
be deeper than usually assumed. Given the relative simplicity of the quantum
turbulence, this may provide new insights into the nature of turbulence.

\begin{acknowledgments}
This work was carried out under the HPC-EUROPA2 project
228398, with the support of the European Community
(Research Infrastructure Action of the FP7). CFB, YAS and AS acknowledge the
support of the Leverhulme Trust (Grants F/00125/AH and RPG-097). We are grateful
to P.A.~Davidson and Y.~Kaneda for discussions and encouragements.
\end{acknowledgments}

\end{document}